\documentclass[prd,twocolumn,showpacs]{revtex4}
\usepackage{hyperref,amssymb,amsmath,mathrsfs,bm,graphicx}
\usepackage{enumitem}
\usepackage{mathrsfs}
\usepackage{color}
\begin{document}
\title{Global energy conservation in nonlinear spherical characteristic evolutions}
\author{{W. Barreto}}
\affiliation{Centro de F\'\i sica Fundamental, Facultad de Ciencias, Universidad de Los Andes, M\'erida, Venezuela}
\begin{abstract}
Associated to the unique 4--parametric subgroup of translations, normal to the Bondi--Metzner--Sachs group, there exists a generator of the temporal translation  asymptotic symmetry. {Such a descriptor of the motion along the conformal orbit near null infinity is propagated to finite regions. This allows us to observe the global energy conservation even in extreme situations near the critical behavior of the massless scalar field collapse in spherical symmetry.}  
\end{abstract}
\date{\today}
\maketitle

\section{Introduction}
As it is well known, in general relativity the condition $T^\mu_{\nu;\mu}=0$ may be interpreted as expressing local conservation of the energy momentum of matter. It turns out that this condition does not, in general, lead to a global conservation of energy law \cite{w84}, that is, a law that states that the total energy content of matter (as an integral involving $T_{\mu\nu}$ over a spacelike hypersurface) is conserved. An exception occurs if a Killing vector field $\xi^a$ is present in spacetime.

Any conservation law should count the gravitational field energy, but it is not possible to construct a tensor expressed only through the metric and their first derivatives (in accordance with the equivalence principle) \cite{h14}. In other words, there is no known meaningful notion of energy density of the gravitational field in general relativity.

Pointing to global issues, Komar \cite{k59} introduced a judicious notion of total energy. An asymptotically flat spacetime (an idealized isolated system) is analogous to a particle in special relativity. The time translation symmetry and the asymptotic treatment near infinity lead to a better notion of total mass in all stationary asymptotically flat spacetimes that are vacuum near infinity.
 
Insightfully, Bondi {\it et al.}, \cite{bvm62}, \cite{s62},
\cite{s62b} found two important results. Namely, that a gravitationally radiating system must lose mass and the identification of the Bondi--Metzner--Sachs (BMS) asymptotic symmetry group. {The asymptotic treatment near future null infinity ($\mathscr{I}^+$) in general relativity was established since then to identify what classical gravitational radiation is, particularly the notion of the news, that is, the information required to forecast the evolution of the system from the initial condition. There is no radiation without the news and this quantity was clearly connected with the mass loss. But the news can be interpreted as boundary conditions on the world tube at $\mathscr{I}^+$ and can also be related with the BMS group. In principle, one can ``transport" that information, using the field equations, from infinity to the interior region where the source is, and see its structure and motion changes.}  
It was not then clear how important the role of the boundary conditions at null infinity was to the formulated characteristic initial value problem (CIVP). {Other authors reformulated the CIVP using conformal techniques (see Ref. \cite{w84} and references therein).} For instance, Winicour and Tamburino \cite{wt65}, \cite{tw66} reformulated the CIVP to finite regions in terms of a null coordinate system based upon a finite world tube rather than upon null infinity. Then, they used the conformal techniques of Penrose \cite{penrose} to derive again the results of Bondi and coworkers and of Sachs. An interesting consequence of the work of Winicour and Tamburino was the representation of the BMS group known as linkages, which offer an invariant way to label energy and momentum in finite regions. Under the appropriate conditions, one can get the total Bondi mass and the news from the linkages.

These ideas and results flourished once again with the advent of numerical general relativity. They have been applied to develop the Pittsburgh characteristic numerical solvers \cite{winicourLL}. At this time, as far as we know, no global energy conservation test has been reported in the nonlinear and extreme regimes. In the linear regime, we reported the global energy conservation in the context of a three--dimensional evolution of a massless scalar field on a Schwarzschild black hole spacetime background \cite{gbf07}. {We report here the nonlinear regime for the global conservation of energy from a simple model to consider elsewhere other dimensionality.}
\begin{figure}[!ht]
\includegraphics[width=3.5in,height=3.5in]{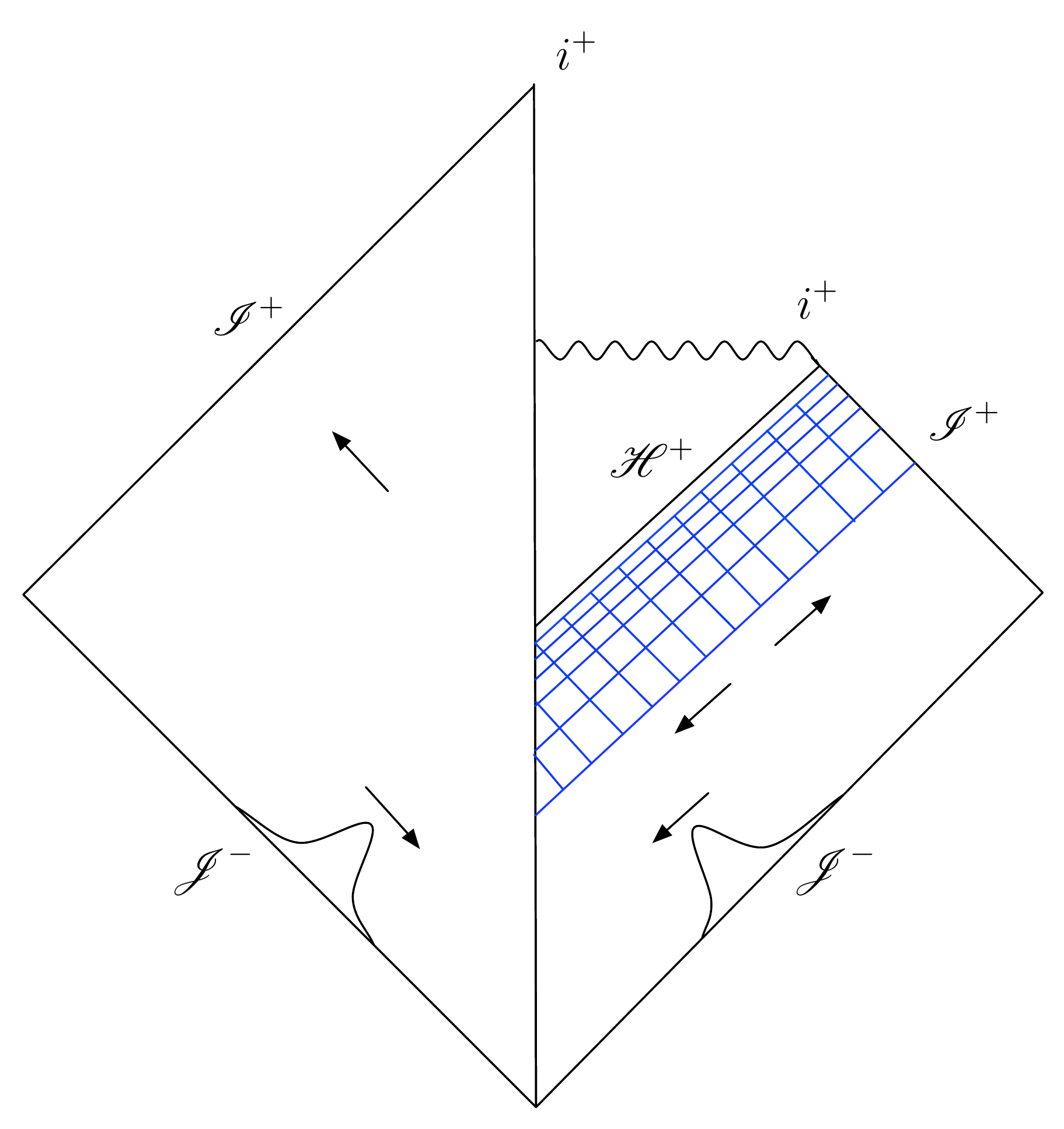}
\caption{Sectors of the initial--boundary value problem for the EKG system. Data for an ingoing pulse is specified on the initial outgoing null hypersurface $\mathscr{J}^-$. A black hole may form (right) or not (left), but always the pulse is scattered to $\mathscr{I}^+$, partially or totally. The final black hole mass depends on the initial amplitude. Evolution slows down near the critical (unstable) solution or in the vicinity of the future event horizon $\mathscr{H}^+$.}
\label{fig:pensorediagram}
\end{figure}
The most simple case to detect and explore ``radiation" is the spherical [one--dimensional (1D)] Einstein--Klein--Gordon (EKG) system (see Fig. 1). In this context, the initial-boundary data determine the fields at $\mathscr{I^+}$, where the spacetime is in fact asymptotically flat. We do not need to specify the news explicitly when we proceed from the Winicour--Tamburino framework, that is, Minkowskian at the center of symmetry ($r=0$),
to march to infinity where the spacetime is not Minkowskian but asymptotically flat.
All the information is encrypted in the initial-boundary data with the unpredictability inherent to a nonlinear evolution, especially near the critical behavior \cite{c93}, \cite{pha05}. Always we can recover Bondi's frame by a simple transformation. {Although complications of gauge considerations are not present at all in the spherical symmetric system considered here, the 1D EKG  will be helpful for nonlinear considerations and future developments in two and three dimensions.} 

In general, out of the linear regime it is desirable to get global energy conservation. The EKG system in spherical symmetry is again a simple and interesting toy model to accomplish that goal. In the linear case, the existence of a temporal Killing vector field and consequently the time independence for the background metric is clear. This has been done successfully for testing a three-dimensional situation \cite{gbf07}. {In the most general spherically symmetric case, we have found the temporal translation generator, as we shall see. Only if we calculate the flux of energy at $\mathscr{I^+}$ (the news) using the appropriate Killing vector there do we get the energy conservation in the Bondi frame, even in extreme situations, that is,
the threshold of black hole formation. This could be relevant because of the inclusion of topological considerations on the notion of total energy. Numerically, this give us confidence due to the possibility of convergence and stability tests in nonlinear regimes.}

In what follows, we revise and report briefly some points of interest related to the aforementioned issues.
\begin{figure}[!ht]
\includegraphics[width=2.3in,height=2.3in]{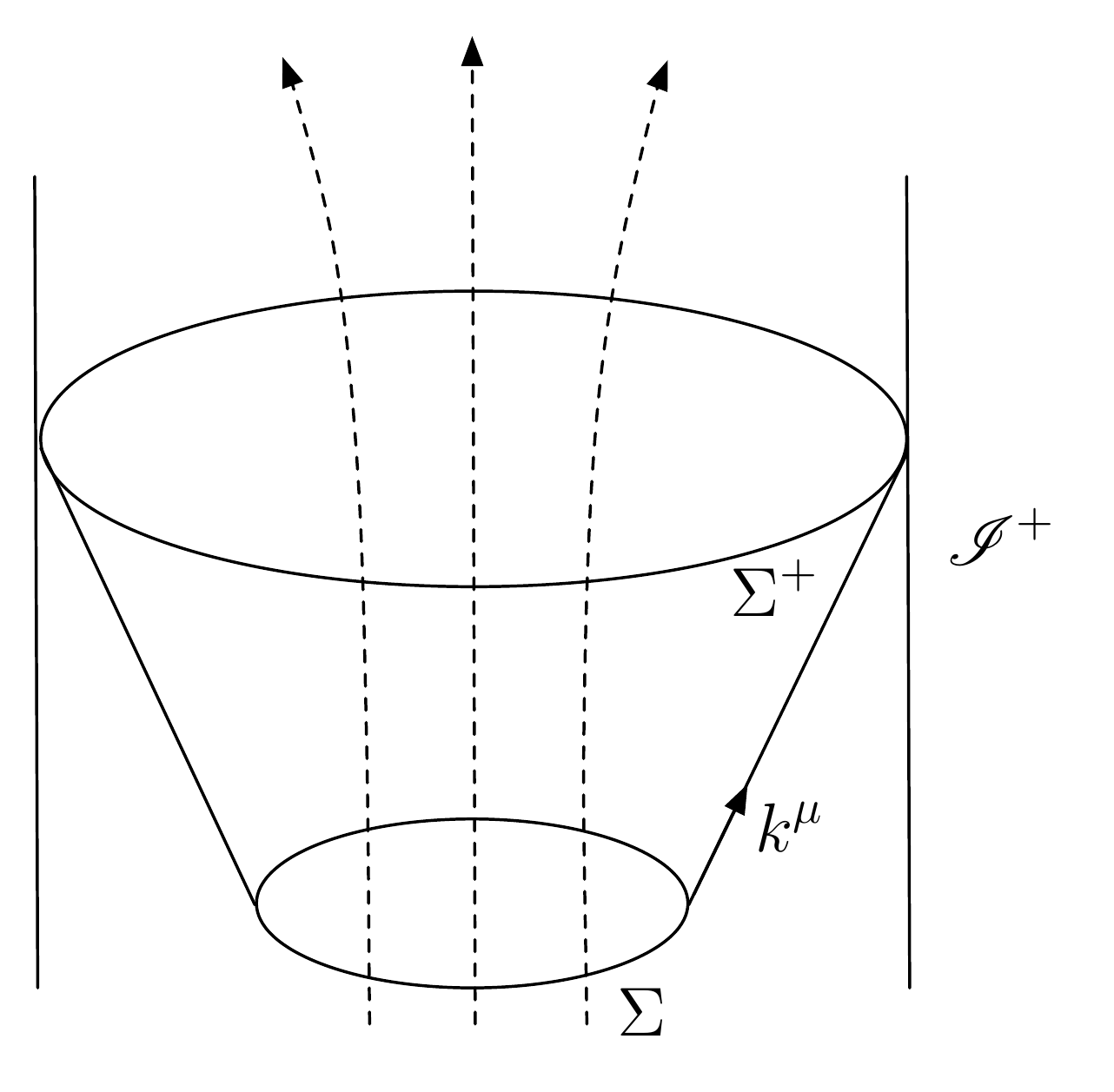}
\caption{Three-dimensional drawing in which $\Sigma$ appears as a closed loop.  $k^\mu$ represent the outgoing  null vector. Dashed lines depict trajectories linking $\Sigma$ and $\Sigma^+$.}
\label{fig:linkage}
\end{figure}
\section{Linkages}
{The Minkowski spacetime has a 10-parametric group of isometries, which leaves the metric invariant, the Poincar\'e group. Four descriptors are associated with translations, and six of them area associated with Lorentz rotations. With $\sigma$ being a spacelike hypersurface, $T^\nu_\mu$ being the energy momentum of some field, and 
$\xi_Q^\mu$ being one descriptor $Q$, then the scalar linear functional \cite{w68}
\begin{equation}
L_Q(\sigma)=\int_\sigma \xi^\mu_Q T^\nu_\mu dS_\nu
\end{equation}
describes the total energy, momentum, and angular momentum of $\sigma$.}

In general, isometries do not exist in curved spacetimes.
However, in asymptotic spacetimes, there exists the notion of asymptotic symmetry at null infinity (${\mathscr I}$), considered as a regular surface composed by two disjoint regions ${\mathscr I}^+$ (future) and ${\mathscr I}^-$ (past), each one with topology  $S^2\times\mathbb{R}$.

A conformal variety to the physical variety can be constructed 
such that the asymptotic symmetry can be defined in terms of the conformal motion of ${\mathscr I}$ \cite{s62}.
This leads to the  BMS asymptotic symmetry group.
Considering only the symmetries of ${\mathscr I}^+$, they can be expressed as infinitesimal conformal transformations \cite{tw66}, \cite{w68}.
The BMS group is a Lie group, which contains: i) the infinite-dimensional subgroup of super translations, which is a factor to the Lorentz group (isomorphic to the conformal group), and ii) the unique, normal and 4--parametric subgroup associated with the Minkowski spacetime translations. A subgroup normal to BMS isomorphic to the Poincar\'e group does not exist \cite{w84}.
Neither does a geometrically intrinsic way to propagate the descriptors of ${\mathscr I}^+$ throughout the spacetime exist. However, a two-surface $\Sigma$ does determine a geometrical prescription for uniquely propagating descriptors from ${\mathscr I}^+$ to $\Sigma$ (see Fig. 2). Each point on $\Sigma$ determines geometrically two null directions that are normal to the local 2-space. All the null outgoing directions on $\Sigma$ define the null hypersurface, which emerges from $\Sigma$ and intercepts ${\mathscr I}^+$ on the 2-space $\Sigma^+$. If $k^\mu$ is the vectorial field normal to this null hypersurface, then the propagation equation 
\begin{equation}
\xi^{(\mu;\nu)}k_\nu = \frac{1}{2}\xi^\beta_{;\beta} k^\mu \label{pe}
\end{equation}
uniquely determines $\xi^\mu$ on the null hypersurface in terms of its value at $\Sigma^+$. 

Now it is possible to define geometrically the linkage $\xi$ throughout $\Sigma$ for each asymptotic descriptor
\begin{equation}
{\cal C}=\int \xi^\nu T^\mu_\nu  d\Sigma_\mu.
\label{linkage}
\end{equation}
In particular, the total Linkage of energy corresponds to a temporal translation. {The linear functional is called now the energy linkage through $\Sigma$ to avoid the specification to any spacelike 3-volume in which we could say the energy resides \cite{wt65}. Topologically, the linkage refers, too, to the four-dimensional analog of the three-dimensional concept of a trajectory passing through a closed loop.} {We shall see an example in which we obtain the global conservation of energy for a general curved spacetime,  in which a black hole may form or not, in the context of spherical symmetry.}
\section{1D EKG}
Using radiation coordinates in spherical symmetry
\begin{equation}
ds^2=e^{2\beta}(Vr^{-1}du^2+2dudr)-r^2(d\theta^2+\sin^2\theta d\phi^2),
\end{equation}
with $V=V(u,r)$ and $\beta=\beta(u,r)$, the EKG system adopts the simple form
\begin{equation}
\beta_{,r}=2\pi r \Phi^2_{,r},
\label{ekg_a}
\end{equation}
\begin{equation}
V_{,r}=e^{2\beta},
\label{ekg_b}
\end{equation}
\begin{equation}
2(r\Phi)_{,ur}-[Vr^{-1}(r\Phi)_{,r}]_{,r}=-\left(\frac{V}{r}\right)_{,r}\Phi,
\label{ekg_c}
\end{equation}
for a massless scalar field, $\Phi=\Phi(u,r)$ minimally coupled to gravitation. The comma indicates a partial derivative. 

Assuming that the scalar field can be expanded in powers of $1/r$ near ${\mathscr{I}}^+$ \cite{gw92},
\begin{equation}
\Phi(u,r)=\frac{Q(u)}{r}+ \frac{c_{NP}}{r^2} + O(r^{-3}),
\end{equation}
the hypersurface equations (\ref{ekg_a}) and (\ref{ekg_b}) lead us to
\begin{equation}
\beta(u,r)=H(u)-\frac{\pi Q^2(u)}{r^2}+ O(r^{-3}),
\end{equation}
\begin{equation}
V(u,r)=e^{2H}\left( r-2M(u)+\frac{\pi Q^2(u)}{r}\right) + O(r^{-3}),
\end{equation}
where $H(u)$ and $M(u)$ are integration functions with physical interpretation. $H$ indicates redshift since Bondi time $u_B$ is related to proper time at the center via the relation $du_B=e^{2H}du$. $M$ is the Bondi mass, which is conserved in this context, as we shall see.
The evolution equation (\ref{ekg_c}) leads to ${c_{NP}}_{,u}=O(r^{-3})$. $c_{NP}$ is the Newman--Penrose constant for the massless scalar field \cite{gw92}.

From the mass of Bondi in the asymptotic form 
\begin{equation}
M=\frac{1}{2}e^{-2\beta}r^2(V/r)_{,r}|_\infty,
\end{equation}
it can be easily shown that
\begin{equation}
M=2\pi\int_0^\infty rVe^{-2\beta}\Phi_{,r}^2 dr.
\label{bm}
\end{equation}
\begin{figure}[!ht]
\includegraphics[width=3.in,height=4.in]{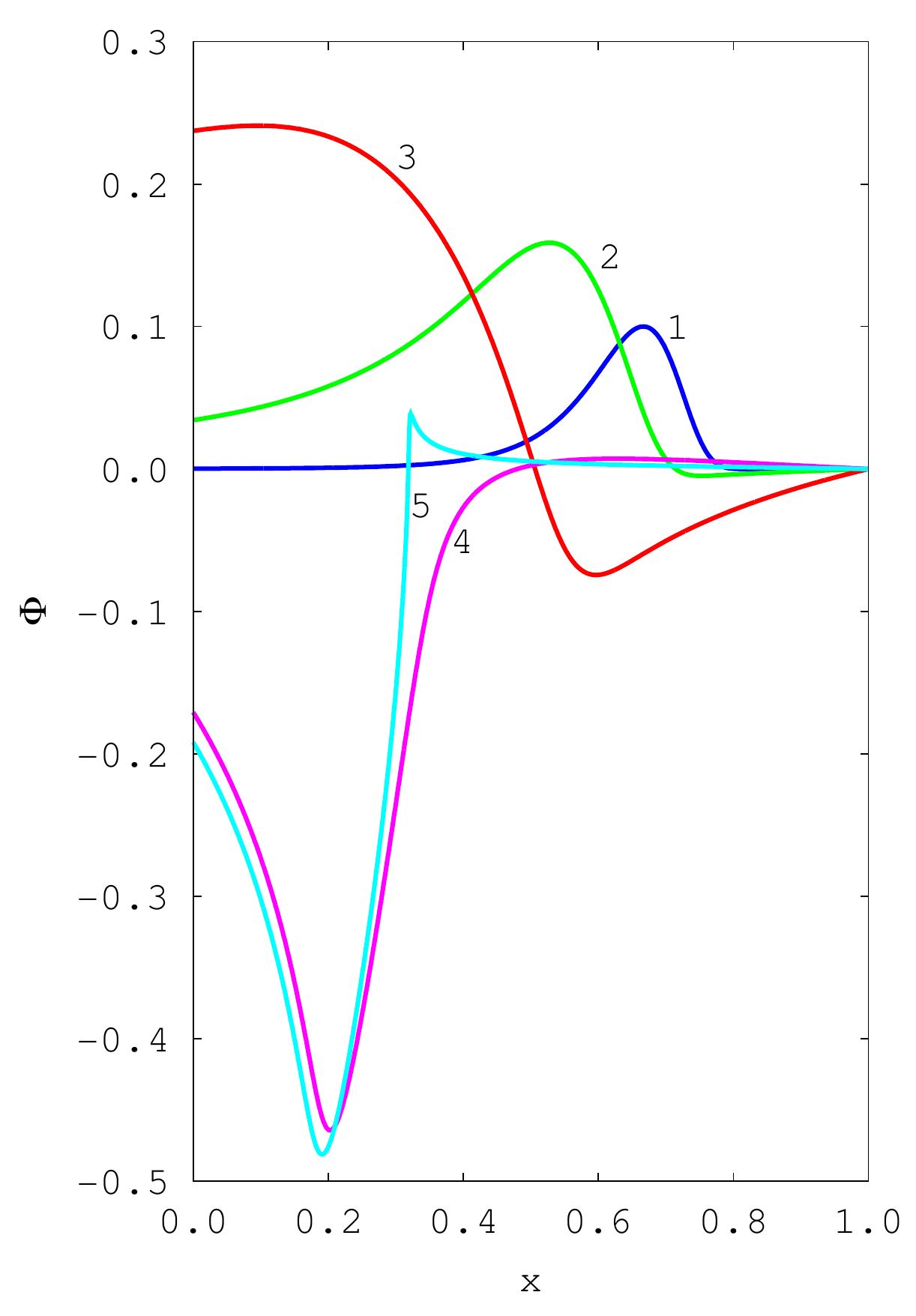}
\caption{Sequence of $\Phi$ as a function of $x$ up to the black hole formation: $u_B\approx 0.0$ (1); $2.4$ (2); $5.7$ (3); $12.4$ (4); $17,7$ (5).}
\label{fig:secI}
\end{figure}
\begin{figure}[!ht]
\includegraphics[width=3.in,height=4.in]{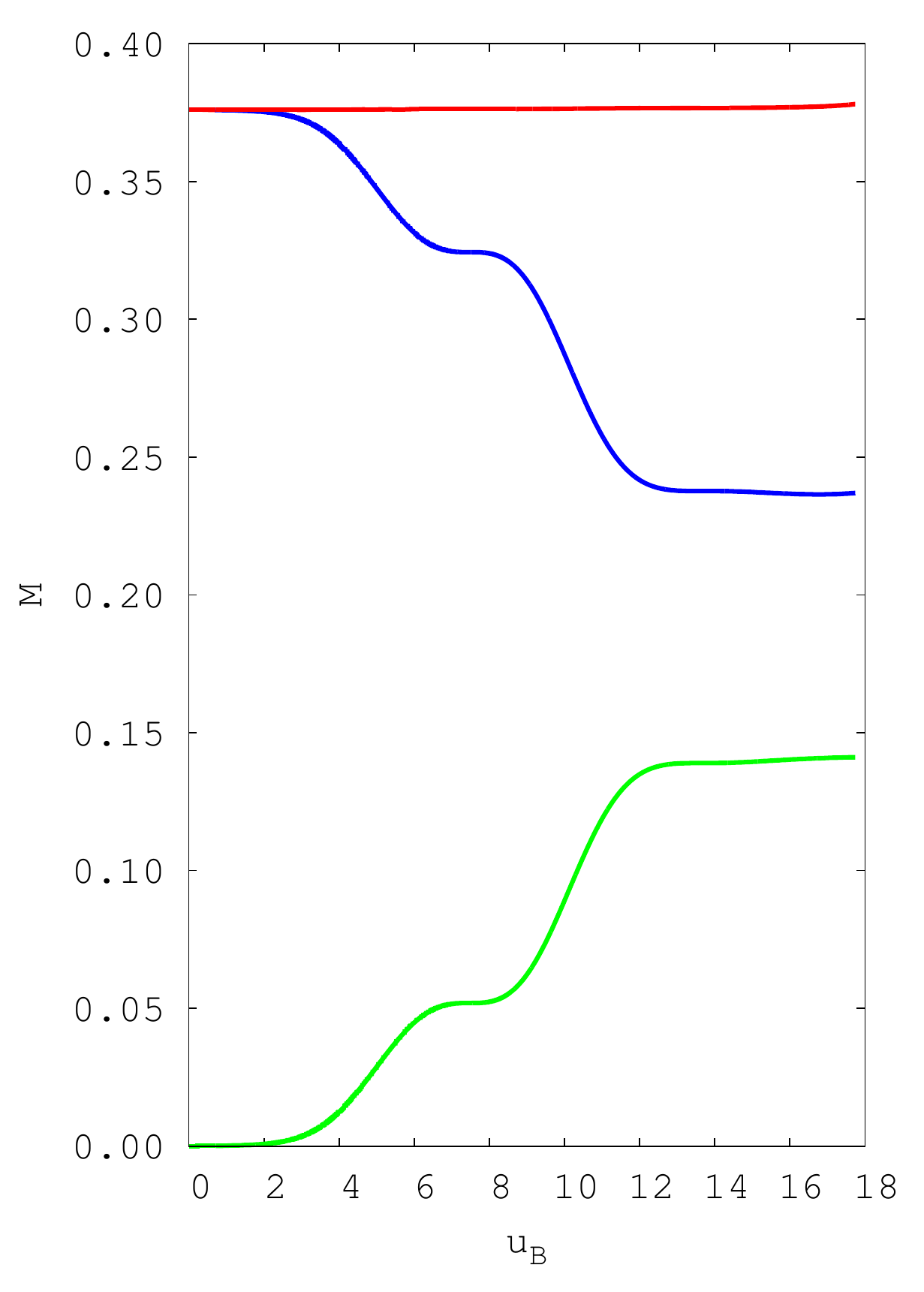}
\caption{Energy conservation as a function of the Bondi time, up to the black hole formation, as displayed in Fig. 3. The descending curve corresponds to the Bondi mass given by Eq. (\ref{bm}); the ascending curve corresponds to the energy flow to infinity calculated as $M_{out}=-\int P du_B$, using Eq. (\ref{powerscri}). The horizontal curve  corresponds to the algebraic conserved sum of both curves.}
\label{fig:energy}
\end{figure}
\begin{figure}[!ht]
\includegraphics[width=3.in,height=4.in]{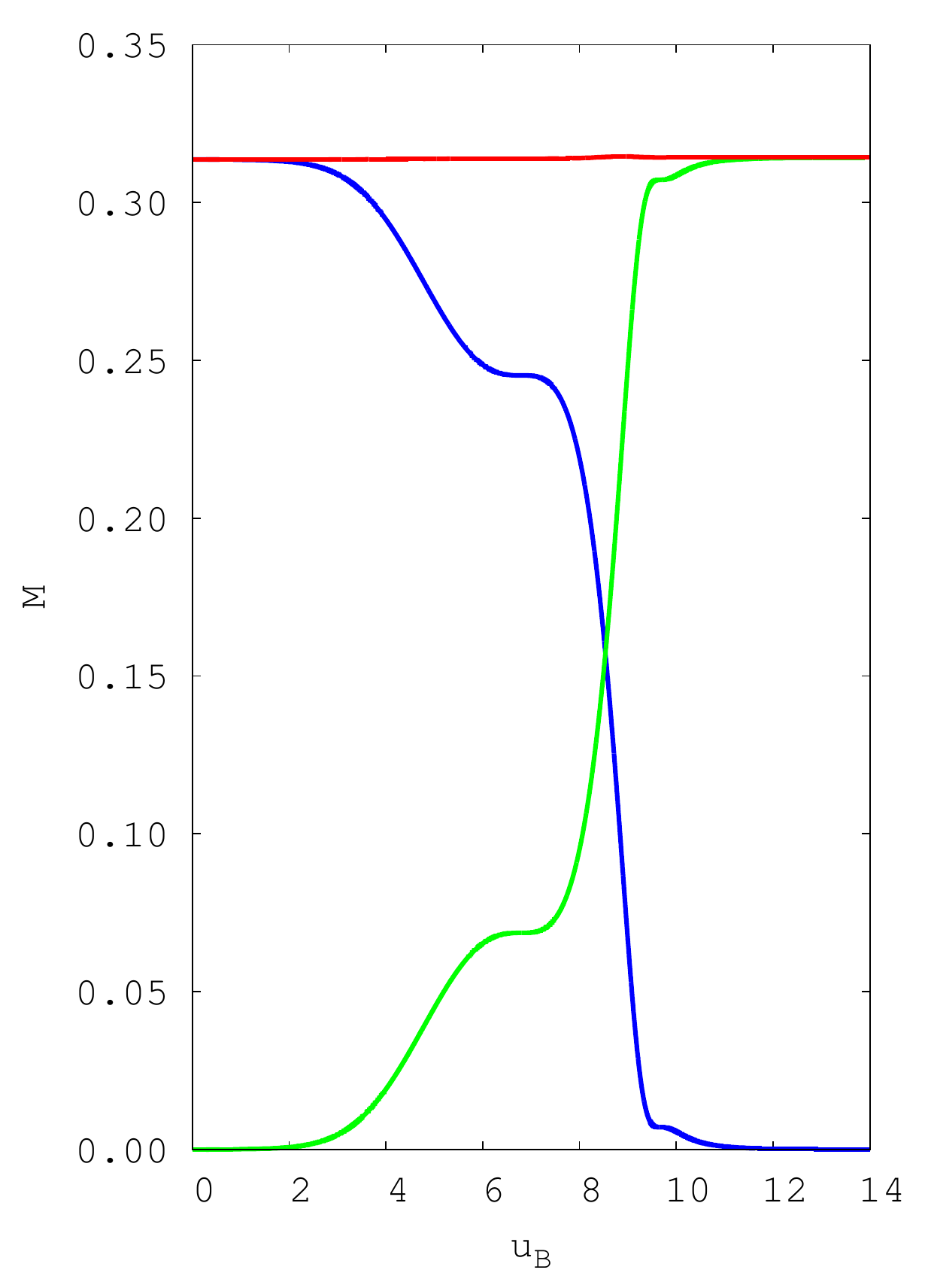}
\caption{Energy conservation as a function of the Bondi time. The evolution is subcritical and mostly nonlinear. The descending curve corresponds to the Bondi mass given by Eq. (\ref{bm}); the ascending curve corresponds to the energy flow to infinity calculated as $M_{out}=-\int P du_B$,  using Eq. (\ref{powerscri}). The horizontal curve corresponds to the algebraic conserved sum of both curves.}
\label{fig:energyII}
\end{figure}
\begin{figure}[!ht]
\includegraphics[width=3.in,height=4.in]{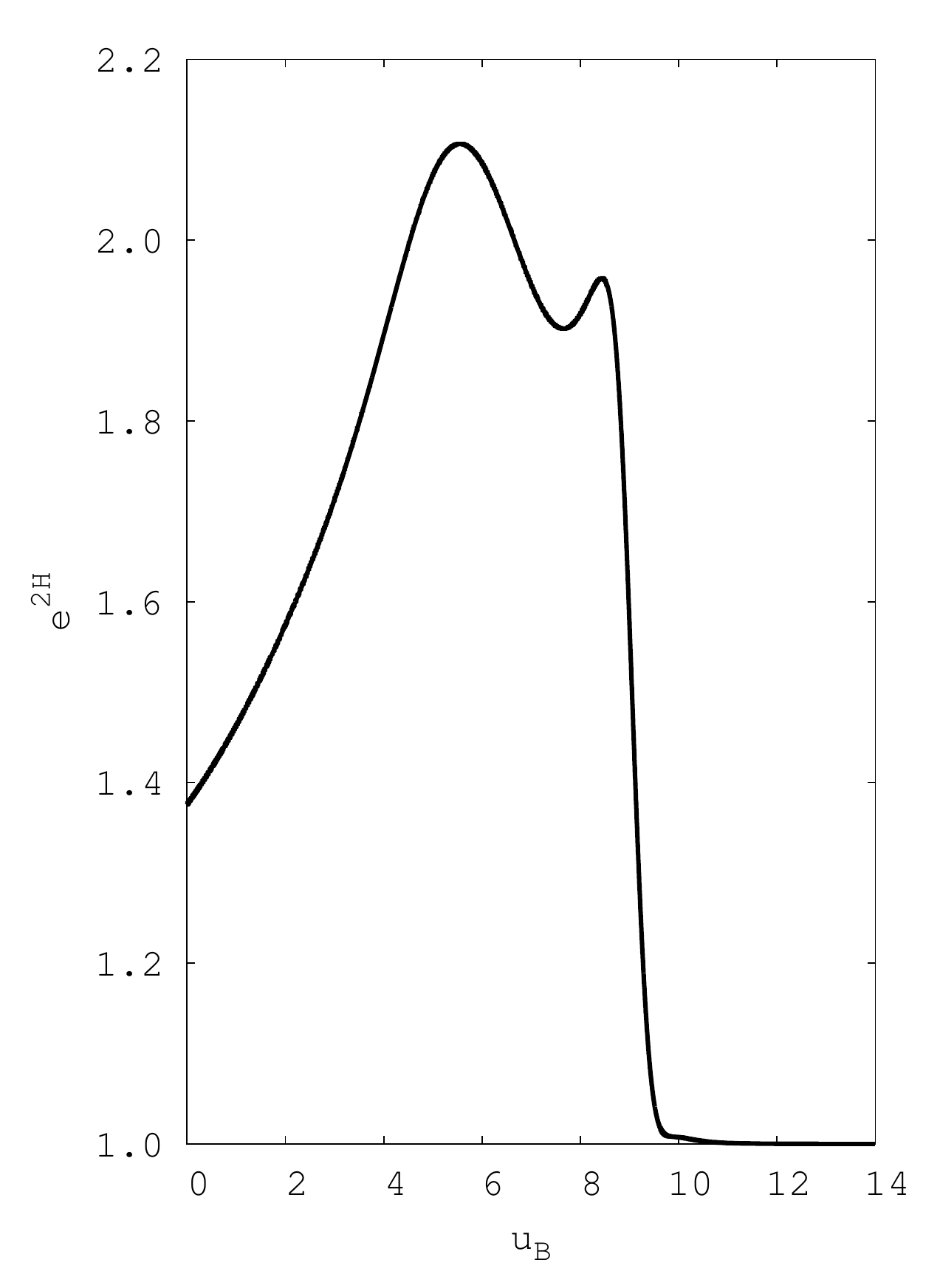}
\caption{Evolution of $e^{2H}$ as a function of the Bondi time for the subcritical case.
It is apparent that the evolution goes from the nonlinear regime to the
linear regime.}
\label{fig:H}
\end{figure}
\begin{figure}[!ht]
\includegraphics[width=3.in,height=4.in]{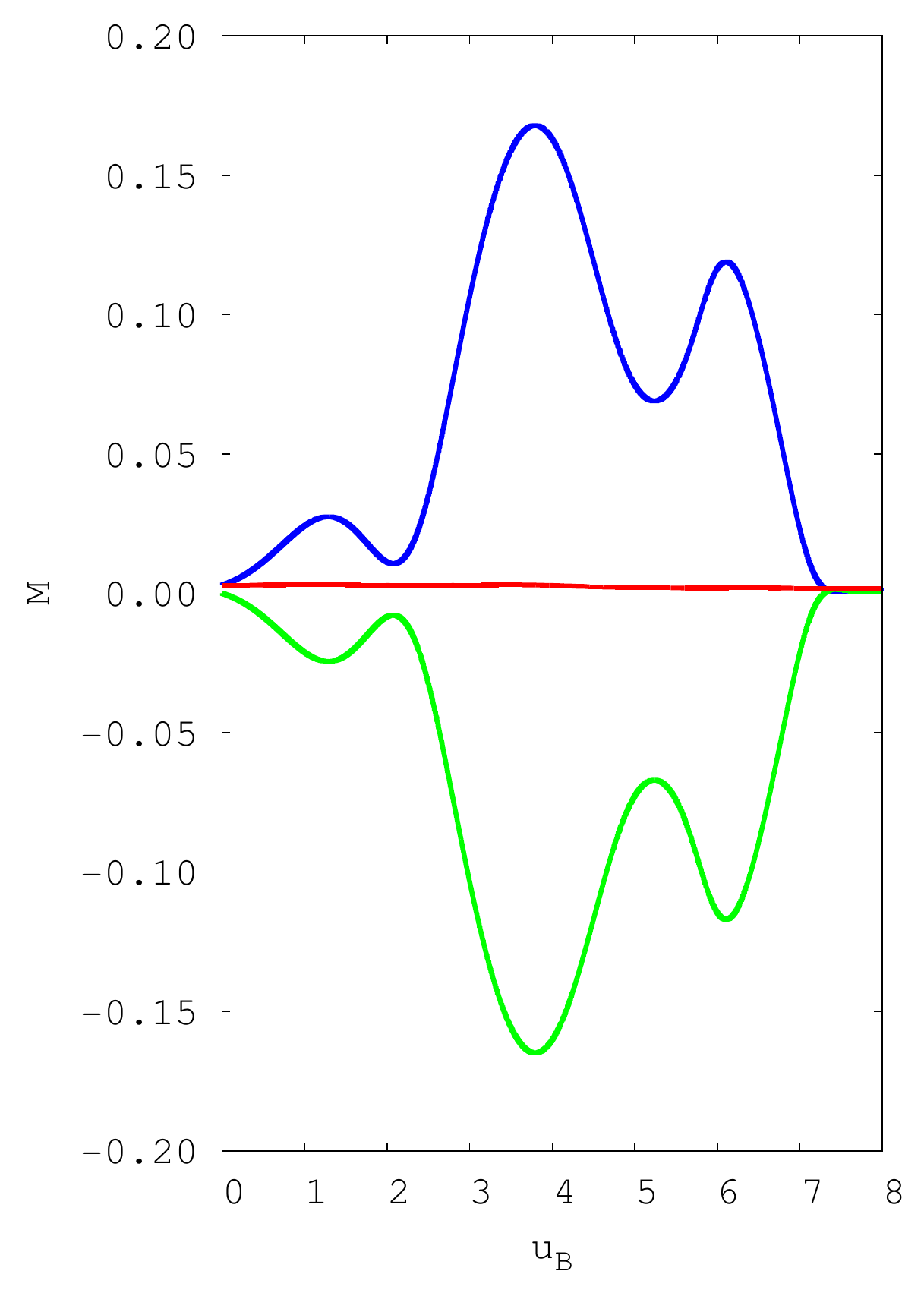}
\caption{Energy conservation as a function of the Bondi time, in a subcritical and nonlinear evolution. The upper curve corresponds to energy given by Eq. (\ref{bm}), but with limits between $r_{in}=1/3$ and $r_{out}=1/2$; the lower curve corresponds to the energy flow calculated as $M_{in}=\int P(u_B,r_{in}) du_B$, $M_{out}=-\int P(u_B,r_{out}) du_B$,  using Eq. (\ref{power}). The horizontal curve corresponds to the algebraic conserved sum $M + M_{out}-M_{in}=$ constant.} 
\label{fig:figure7}
\end{figure}

To compare with the invariant (\ref{linkage}), we have to take into account
that 
\begin{equation}
T^u_u=e^{-2\beta}\frac{V}{2r}\Phi_{,r}^2
\end{equation}
and
\begin{equation}
d\Sigma_u=e^{2\beta}r^2dr \sin\theta d\theta d\phi.
\end{equation}
{Therefore, the temporal translation generator is given by
\begin{equation}
\xi^\nu=e^{-2\beta}\delta^\nu_u,
\label{tk}
\end{equation} 
which satisfies $\xi^\nu_{;\nu}=0$ and therefore represents a global isometry that leads us to the energy conservation.}

We use the generator (\ref{tk}) to determine from Eq. (\ref{linkage}) the power radiated,
calculated first as the flux across the surface $r=$ constant 
\begin{equation}
P=\int T^r_u e^{-2\beta}d\Sigma_r,
\end{equation}
where 
\begin{equation}
T^r_u=e^{-2\beta}(\Phi_{,u} -Vr^{-1}\Phi_{,r})\Phi_{,u},
\end{equation}
and
\begin{equation}
d\Sigma_r=r^2\sin\theta d\theta d\phi.
\end{equation}
Thus,
\begin{equation}
P=-4\pi r^2 e^{-4\beta}(\Phi_{,u} -Vr^{-1}\Phi_{,r})\Phi_{,u}
\label{power}
\end{equation}
can be written as the power radiated to infinity,
\begin{equation}
P=-4\pi e^{-4H} [(r\Phi)^2_{,u}]_{{\mathscr I}^+},
\label{powerscri}
\end{equation}
from which we obtain a version of the Bondi's formula
to to connect the news with mass loss:
\begin{equation}
\frac{dM}{du_B}=-4\pi \left[\frac{dQ}{du_B}\right]^2.
\end{equation}

Note that the power radiated is expressed as the variation of the mass with respect to the Bondi time, which is crucial to obtain the balance of energy. 
The expressions for the total energy and power radiated have been reported and used for numerical purposes \cite{gw92,giw92,bglw96,l00,sfp02,pha05}, \cite{gbf07}, but as far we know they had not been used in extreme situations of gravitational collapse. 

We used an extended version of the characteristic Pittsburgh code \cite{bglw96} to reproduce the
critical collapse of a massless scalar field minimally coupled to gravitation. 
For this purpose, we consider near $r=0$ that
\begin{equation}
\Phi=\Phi_0(u)+\Phi_1(u) r + \Phi_2(u) r^2 + O(r^{3}).
\end{equation}
The generality of this expansion toward the origin is validated by the analytical
and numerical results, obtained by other authors \cite{c93}, \cite{g95}, \cite{kha95}, \cite{hs96} and reproduced by us. It remains to mention here the use of the compactified radial coordinate,
$$x=\frac{r}{1+r},$$
so that points at $\mathscr{I}^+$ are included in the numerical grid at $x = 1$.
%%%%%%%%%%%%%%%%%%%%%%%%%%%%%%%
\section{Evolutions}
%%%%%%%%%%%%%%%%%%%%%%%%%%%%%%%
{Figure \ref{fig:secI} shows the scalar field evolution for the initial datum 
\begin{equation}
\Phi(u=0,r)=\lambda e^{-(r-r_0)^2/\sigma^2}
\end{equation}
with  $r_0=2$, $\lambda=10^{-1}$, $\sigma=0.8$. The radial grid in the compactified
coordinate is $N_x=10^3$. In this case the collapse is supercritical and consequently forms a black hole. When the redshift is about $10^4$, the evolution is stopped.
Figure \ref{fig:energy} displays the energy conservation for the same initial datum, with $0.6\%$ of maximum variation relative to the initial energy. From Figs. 3 and 4 we observe that the residual energy corresponds to $r_{\mathscr{H}}=2M_{\mathscr{H}}\approx 0.47$; about $37\%$ of the energy is radiated to infinity. Figure \ref{fig:energyII} shows the energy conservation for the same initial datum, but now with $\lambda=9\times 10^{-2}$. In this subcritical case, initially off the linear regime, the energy is completely radiated to infinity. Figure \ref{fig:H} shows the redshift at ${\mathscr I}^+$, from the nonlinear regime to the linear regime.}

{We want to illustrate how the power radiated (\ref{power}) can be used in finite regions. For example, let us determine the energy content between $r_{in}=1/3$ and $r_{out}=1/2$ and the energy flow throughout the surfaces defined by these radii. Figure 7 displays such a computation for the same conditions of Figs. 5 and 6. 
Now, the Bondi time lapse has to be calculated using $\beta$ at $r_{in}$ or $r_{out}$; otherwise, the energy balance is not reliable. In this specific case, we have used the Bondi time at $r_{in}$. 
Initially, the considered volume contains a small quantity of energy that is increased to a maximum of about $u_B\approx3.8$. The algebraic sum of the two fluxes at $r_{in}$ and $r_{out}$ allow us to observe the energy conservation.}
%%%%%%%%%%%%%%%%%%%%%%%%%
\section{Conclusions}
%%%%%%%%%%%%%%%%%%%%%%%%%
{The asymptotic descriptor of the temporal translation associated to the unique 4-parametric subgroup  of translations, normal to the BMS group, was identified in the context of spherical symmetry. Such a descriptor of the motion along the conformal orbit near null infinity is propagated to finite regions. Using the energy linkage for a massless scalar field minimally coupled to gravitation, we obtain the global energy conservation when a black hole forms and when the scalar field disperses completely to infinity. Moreover, to illustrate the balance of energy at finite regions, we make a calculation between two experimental spheres with satisfactory results. In this way, the global energy conservation for any regime of the 1D EKG was computed.} 

{If we compare developments in the Arnowitt-Deser-Misner (ADM) $3+1$ formulation with the characteristic approach in this respect, we shall see that the energy (mass) in ADM has been calculated, but its global conservation has not been verified. It has been limited to establish its positivity and monotonic decreasing in time (see Ref. \cite{w84} and references therein). We may now verify the global energy conservation in extreme scenarios when using the ADM $3+1$ formulation. This is possible modulo the Bondi--Sachs gauge \cite{fg07}, at least in the spherical symmetry EKG system. Our main interest is to make an extension to less symmetric spacetimes, where gravitational and electromagnetic radiation could be present.} 
Some analogous and extended work is in progress for the 2D EKG system in the characteristic (ingoing) formulation.  

\section*{ACKNOWLEDGMENTS}
Thanks to L. Rosales and C. Peralta for their valuable comments.
 
\end{document}